\documentstyle[12pt]{article}

\begin{document}
\def\la{\mathrel{\mathpalette\fun <}}
\def\ga{\mathrel{\mathpalette\fun >}}
\def\fun#1#2{\lower3.6pt\vbox{\baselineskip0pt\lineskip.9pt
        \ialign{$\mathsurround=0pt#1\hfill##\hfil$\crcr#2\crcr\sim\crcr}}}
\newcommand {\eegg}{e^+e^-\gamma\gamma~+\not \! \!{E}_T}
\newcommand {\mumugg}{\mu^+\mu^-\gamma\gamma~+\not \! \!{E}_T}
\renewcommand{\thefootnote}{\fnsymbol{footnote}}
\bibliographystyle{unsrt}

\begin{flushright} \small{
UAB-FT-400\\
UMD-PP-97-44\\
Fermilab-PUB-96/373-A\\
hep-ph/9610458\\
October, 1996}
\end{flushright}

\vspace{2mm}

\begin{center}

{\Large \bf Astrophysical  bounds on superlight gravitinos}

\vspace{1cm}

{\bf J. A. Grifols$^{{\rm a)}}$, R. Mohapatra$^{{\rm b)}}$ and
A. Riotto$^{{\rm c)}}$}
 \vspace{0.3cm}

{\em a)  Grup de F\'{\i}sica Te\`orica and IFAE,
Universitat Aut\`onoma de Barcelona,\\
08193 Bellaterra,  Spain\\

b) Department of Physics, University of Maryland, College Park, Md-20742\\

c)  Fermilab
National Accelerator Laboratory, Batavia, 
Illinois~~60510-0500, USA              }

\end{center}

\vspace{2mm}
\begin{abstract}

We derive the allowed mass range for the superlight gravitino
present in a large class of supersymmetric models from
the observed neutrino luminosity from Supernova 1987A. 
We find that for photino masses of order of 100 GeV, 
the mass range $2.6\times 10^{-8}$ eV $\leq
 m_{\tilde{g}}\leq 2.2\times 10^{-6}$ eV for the gravitino $\tilde{g}$
is excluded by SN1987A observations. 
Unlike the bounds on $m_{\tilde{g}}$
from nucleosynthesis, the bounds in the present paper do not depend 
critically upon the uncertainities of the observational input.

\end{abstract}
 
\newpage
\renewcommand{\thefootnote}{\arabic{footnote})}
\setcounter{footnote}{0}
\addtocounter{page}{-1}
\baselineskip=24pt

In models of dynamical supersymmetry breaking where gauge interactions 
mediate the breakdown of supersymmetry \cite{dine} the gravitino is naturally 
light ($\ll$1 TeV). Indeed, it is
supposed to be the lightest supersymmetric particle (LSP). 
While the precise value of its mass depends on the details of any
given model, the general expectation is that its order of magnitude
is given by $m_{\tilde{g}}\simeq \Lambda^2/M_{P\ell}$ where $\Lambda$
is a typical SUSY breaking scale. For $\Lambda\simeq 100$ GeV,
one gets $m_{\tilde{g}}\simeq 10^{-6}$ eV or so.
It is however possible to imagine more general relations between
the gravitino mass and the $\Lambda$ such as $\Lambda\simeq
(m_{\tilde{g}}/M_{P\ell})^{1-2/3q} M_{P\ell}$ \cite{ellis}, in which case
scenarios with even lighter gravitinos could emerge (e.g. when
$q> 3/4$). More importantly, 
the coupling of the gravitino to matter is inversely proportional
to its mass (i.e. 
$g_{ff\tilde{g}}\simeq M^2_{\tilde{f}}/(m_{\tilde{g}}M_{P\ell})$) 
which increases as its mass
decreases. Thus information about the gravitino mass provides us 
not only with direct information about the magnitude of supersymmetry
breaking scale and the nature of SUSY breaking 
but also about possible experimental manifestation
of supersymmetry (the lighter it is, the easier it is to produce them). 
Furthermore, cosmological significance of the gravitino also depends
on its mass; for instance, only if its mass is as large a keV,
can it be relevant for the dark matter problem~\cite{borgani}.

In the absence any direct laboratory information 
about the gravitino mass\footnote{We however note that 
recent interpretation of the CDF
$\gamma\gamma e^+e^-$ event in terms of a light gravitino decay\cite{kane}
of the photino seems to imply a gravitino mass in the eV range.},
it is necessary to look at phenomena that provide indirect information
on $m_{\tilde{g}}$. Astrophysics and cosmology are the obvious places.
In the domain of cosmology, the first place where gravitinos could
alter the conventional arguments is the era of Big Bang 
Nucleosynthesis (BBN). It has been argued that superlight gravitino 
production can add to
the energy density of the universe and thereby effect the predictions for
the primordial $^4$He abundance unless the gravitino mass 
is below some $10^{-6}$ eV~\cite{gherghetta}. 
This result follows from a careful reexamination of previous work
of Moroi et al. ~\cite{moroi}. 
This bound relies on allowing for less than 0.6 effective 
extra light neutrino degrees of freedom in the usual primordial $^4$He 
abundance calculation. However, concern about 
systematic uncertainty in the $^4$He abundance as well as the
 chemical evolution of $^3$He has led to the  re-examination  
this important limit \cite{copi} with the result that no more   
than the equivalent of 4 massless neutrino species are allowed. 
Here, we want to point out that, should one use 
this more conservative bound $\Delta N_\nu \leq 1$ \cite{copi,sarkar}, and 
repeating the reasoning that led to $m_{\tilde{g}} \geq 10^{-6}$ eV, one 
would reach the much less restrictive constraint $m_{\tilde{g}} \geq 3.2 
\times 10^{-9}$ eV. The reason is that if $\Delta N_{\nu}$ is allowed to be one
and there are no other light particle species than the particles of the 
standard model, then the gravitino need not decouple from the thermal
bath of the universe prior to the BBN era. 
Its coupling to neutrinos and photons and hence its mass is less 
restricted\footnote{Note however that if $m_{\tilde{g}}$ is less than
$10^{-8}$ eV or so,the amplitude for $e^+e^-\to \tilde g\tilde g$ becomes
of order one which would be inconsistent with laboratory observations.}.
Thus, the cosmological bound on the gravitino 
mass depends very dramatically on how many extra equivalent massless 
neutrinos does BBN actually allow. This is a not very satisfactory 
situation and one would prefer a bound less dependent on the actual 
number of extra light neutrino species permitted. Indeed, one would 
like to have an independent assessment of the bound on the gravitino 
mass coming from another physical input. In this letter, 
we shall use the observed SN1987A signal to limit the mass of the gravitino.

Since the gravitino has large coupling to the particles present
in the supernova such as the photons and the leptons, for certain
mass range of the gravitino, we expect significant gravitino production
in the supernova core. If gravitino coupling is in such a range that,
its mean free path after production exceeds the supernova radius of
$10-30$ Km, then it will escape carrying energy from the supernova.
The luminosity associated with gravitino emission by the core of the 
protoneutron star must however be bounded by $L\leq 10^{52}$ erg/s.  
This bound in fact applies to any particle other than neutrinos 
and is the well known constraint 
inferred from the detected neutrino events by Kamiokande and IMB on 
february 1987~\cite{bionta} and from the predictions 
of models of stellar collapse \cite{burrows}. 

The mechanisms that, in principle, contribute to gravitino production 
in the hot SN core are: gravitino pair production by photon-photon 
collisions, nucleon-nucleon bremsstrahlung of gravitino pairs (i.e. 
$NN \to NN \tilde g \tilde g)$, bremsstrahlung of gravitinos in 
electron-electron scattering (i.e. $e^-e^- \to e^-e^- \tilde{g} \tilde{g})$, 
and pair production in $e^+e^-$ annihilation. All of these processes 
can be evaluated using the effective Lagrangian obtained from the $N=1$ 
supergravity theory which is explicitly given in ref. 
\cite{gherghetta}. The leading 
contribution comes from $\gamma \gamma \to \tilde g \tilde g$. Other 
processes are subdominant.

Let us first discuss nucleon-nucleon bremsstrahlung of the $\tilde g \tilde g$
pair. Since the nucleon is nonrelativistic, we can consider the simple
pion exchange model for the nucleon-nucleon scattering part \cite{fr} of
the relevant Feynman diagram. There are four diagrams where the gravitino
pairs could be emitted from the external legs. They all cancel in the 
nonrelativistic limit \cite{fr}. They could also be emitted from the
internal pion line. To see why it is suppressed for the gravitinos 
pair emission can be seen as follows. The effective four-Fermi interaction
involving the quark-anti-quark and $\tilde g\tilde g$ is parity conserving
in the limit that left and right-handed squarks are degenerate in mass.
Since the gravitino is a Majorana particle, the effective coupling
can be inferred from ref. \cite{gherghetta} to be of pure axial
vector type, i.e. $\bar{q}\gamma^{\mu}\gamma_5 q \bar{\tilde g}\gamma_{\mu}
\gamma_5\tilde g$. Since the axial vector current has odd G-parity,
its matrix element between pion states vanishes. It is worth noting that while
strict degeneracy between the squark states is not phenomenologically required,
the fact that atomic parity violation experiments agree so well with
the predictions of the standard model implies that they are degenerate
at least to less than a few percent resulting in the suppression as mentioned.
There is further suppression compared to $\gamma\gamma \to \tilde g\tilde g $ 
coming from the fact that the number density of 
photons at $T \sim 50$ MeV  is larger then 
the number density of nucleon scatterers at (super) nuclear matter 
densities of the protoneutron star. 

Bremsstrahlung production in electron-electron collisions is supressed 
by $\alpha^2$ and, again by the fact that 
$n_{e^-}=n_p\ll n_\gamma$. Since positrons are rare in the stellar 
medium, $e^+e^-$ annihilation is also unimportant.
Therefore, the bulk of the gravitino emissivity is associated to the 
process $\gamma (k_1) + \gamma (k_2) \to \tilde{g}(q_1)+ 
\tilde{g}(q_2)$ whose luminosity is explicitly given by

\begin{eqnarray}
L=V \int \frac{d^3k_1}{2 k^0_1} \frac {d^3 k_2}{2 k_2}
 \frac{d^3 q_1}{2q^0_1}\frac{d^3q_2}{2 q^0_2} \frac{1}{(2 \pi)^{12}}
n_\gamma (k^0_1) n_\gamma (k^0_2) \mid {\cal M}\mid^2 (q^0_1+q^0_2) (2 
\pi)^4 \delta^4 (P_f-P_i), 
\end{eqnarray}  
where ${\cal M}$ is the amplitude, $n_\gamma$ is the photon Bose-Einstein 
distribution function and $V$ is the volume of the core.
Using energy conservation and the definition of the cross-section for 
the process $\gamma \gamma \rightarrow \tilde{g} \tilde{g}$, Eq. (1) can be 
recast in the form
\begin{equation}
L = \frac{V}{ (2 \pi)^6} \int d^3k_1 d^3k_2 n_\gamma(k^0_1)n_\gamma 
(k^0_2) \frac {k^0_1+k^0_2}{k^0_1k^0_2}\: k_1 \cdot k_2 \ \  \sigma
(\gamma \gamma \rightarrow\tilde{g} \tilde{g}),  
\end{equation}
where $
\sigma (\gamma \gamma \rightarrow \tilde{g} \tilde{g}) = 
\frac{1}{576 \pi} \left 
( \frac{1}{M_{P\ell}m_{\tilde{g}}} \right)^4 m^2_{\tilde \gamma} s^2$, given 
in refs. \cite{gherghetta,bhattacharya}
and valid in the kinematical domain $s\ll 
m^2_{\tilde{\gamma}}$ ($s$ is the C.M. 
energy squared, and $m_{\tilde{\gamma}}$ is the photino mass).
Because $({\rm e}^{k_0/T}-1)^{-1}> {\rm e}^{-k^0/T}$ always, it follows that
\begin{equation}
L> \frac{V}{(2 \pi)^6} \int d^3k_1 d^3k_2 {\rm e}^{-k^0_1/T} {\rm e}^{-k^0_2/T} 
(k^0_1+k^0_2) (1-\cos \theta) \sigma (\gamma \gamma \to \tilde g \tilde 
g)
\end{equation}
where $\theta$ is  the angle between the vectors  $\vec k_1$ and $\vec k_2$.
The above integral can be easily performed, and the final result is
\begin{equation}
L> \frac{20}{\pi^5} \left ( \frac{1}{M_{P\ell} m_{\tilde{g}}} \right )^4 
m^2_{\tilde \gamma} \ V \ T^{11}. 
\end{equation}
Now we can use the observational bound $L\leq10^{52} $ erg/s and obtain,
\begin{equation}
m_{\tilde{g}}>2.2 \times 10^{-6}\left (\frac{m_{\tilde \gamma}}{100\: 
{\rm GeV}}\right)^{1/2} \left (\frac{T}{50\:
{\rm MeV}}\right)^{11/4} \left (\frac{V}{4.2 \times 10^{18}\: {\rm cm}^3}\right)^{1/4}\:{\rm eV}.
\end{equation}
Of course, this bound is meaningful only if gravitinos free-stream out 
of the star. Therefore, we should check whether this is indeed true for 
gravitino masses that exceed $2.2 \times 10^{-6} $ eV.
The main opacity source comes from elastic scattering of the produced 
gravitinos with photons in the stellar plasma, $\tilde g \gamma 
\to \tilde g \gamma$, 
because {\it i)} $\sigma (\tilde g \gamma \to \tilde g \gamma)\gg 
\sigma (\tilde g f \to \tilde g f)$ where   $f$ denotes eiher  electrons or 
nucleons, and 
{\it ii)} $n_\gamma\gg n_f$. Hence the mean-free-path of gravitinos in the 
stellar core is, 
\begin{equation}
\lambda \sim \frac {1}{n \sigma} \sim \frac{2 \pi^3}{9 \zeta (3)} 
m_{\tilde{g}}^4 M_{P\ell}^4 m_{\tilde \gamma}^{-2} T^{-7} 
\end{equation}
or, putting numbers,
\begin{equation}
\lambda \sim 1.7 \times 10^{4}  \left (\frac{m_{\tilde{g}}}{2.2 \times 
10^{-6}\:  {\rm eV}} \right )^4 \:{\rm Km}\gg 10 \left 
(\frac{R_{{\rm core}}}{10 \ {\rm Km}} \right 
)\:{\rm Km},
\end{equation}
which means that gravitinos, once produced, leave the star without 
rescattering.

It is clear from Eq. (7), on the other hand, that for sufficiently 
small $m_ {\tilde{g}}$, gravitinos would diffuse in the core and therefore the bound 
in Eq. (5) would be no longer valid. If energy is depleted over times 
larger than the typical $\sim 1$ second period of neutrino energy 
emission from the neutrino-sphere, the luminosity associated with 
gravitino emission would be lower and eventually compatible again with 
the observational bound $L< 10^{52} $ erg/s. Let us next estimate the 
range of masses $m_{\tilde{g}}$ allowed by the slow $(t_{{\rm diff}}\geq 1$ sec) 
diffusion of gravitinos.
It takes a time $t=\frac{\lambda}{c} N$ and $N$ scatterings for the 
gravitinos in the core to random-walk over a distance $R_{{\rm core}} \sim 
\lambda \sqrt {N}$ and leave the star. If one requires $t\sim 1$  sec, 
the particle would leave the star after $\sim 9\times 10^8$ collisions. 
This implies a mean-free-path on the order of 0.3 m. Using Eq. (6) we 
then obtain that $t_{{\rm diff}} \geq 1$ sec for
\begin{equation}
m_{\tilde{g}} \leq 2.6 \times 10^{-8} \left ( \frac{m_{\tilde 
\gamma}}{100\ {\rm GeV}}\right )^{1/2} \left (\frac{T}{50\:{\rm  MeV}} \right 
)^{7/4}\:{\rm eV}. 
\end{equation}
SN physics, therefore, forbids the range of gravitino masses from about 
$2.6 \times 10^{-8}$ eV up to $2.2 \times 10^{-6} $ eV. This result 
complements previous work on astrophysical bounds \cite{fukugita}.

In conclusion, we have derived constraints on the mass of a superlight
gravitino from supernova 1987A observations and find an interesting
range of masses excluded by these considerations. In terms of the
generalized mass formula for the gravitinos given in Ref. \cite{ellis},
our results imply $q\leq 3/4$ and the scale of supersymmetry breaking
$\Lambda\geq 100$ GeV.

\bigskip
\noindent{\Large \bf{Acknowledgements}}
\bigskip

Work of J. A. G. is partially supported by the CICYT 
Research Project AEN95-0882 and 
by the Theoretical Astroparticle Network under the EEC Contract No. 
CHRX-CT93-0120 (Direction Generale 12 COMA). The work of R. N. M.
is supported by the National Science Foundation grant no. PHY-9421386
and the work of A. R. is supported by the DOE and NASA under grant no.
NAG5-2788. J.A. Grifols wishes to 
thank Prof. E. Kolb and the Astrophysics Group at FNAL for their kind 
hospitality.



\newpage


\begin{thebibliography}{99}

\bibitem{dine} M. Dine, A.E. Nelson, Y. Nir, and Y. Shirman,
 hep-ph/9507378 {\sl Phys. Rev.} {\bf D53} (1996) 2658; A.E. Nelson, 
hep-ph/9511218;
M. Dine and A.E. Nelson, {\sl Phys. Rev.} {\bf D48} (1993) 1277.
M. Dine, A.E. Nelson, and Y. Shirman, {\sl Phys. Rev.} {\bf D51} (1995) 
1362.

\bibitem{ellis} J. Ellis, K. Enqvist and D. V. Nanopoulos, Phys. Lett. {\bf B
147} (1984) 99.

\bibitem{borgani}S. Borgani, A. Masiero and M. Yamaguchi, 
hep-ph/9605222.

\bibitem{kane} S. Ambrosanio, G. Kane, G. Kribs, S. Martin and S. Mrenna,
Phys. Rev. Lett. {\bf 76} (1996) 3498; S. Dimopoulos, M. Dine, S. Raby and S.
Thomas, {\it ibid} {\bf 76} (1996) 3494.

\bibitem{gherghetta} T. Gherghetta, hep-ph/9607448.

\bibitem{moroi} T. Moroi, H. Murayama, and M. Yamaguchi, {\sl Phys. 
Lett.} {\bf B303} (1993) 289.

\bibitem{copi} C.J. Copi, D.N. Schramm, and M.S. Turner, 
Fermilab-Pub-96/122-A.

\bibitem{sarkar} S. Sarkar, OUTP-95-16P; Rep. Prog. Phys. (to appear);
K. Olive, Invited talk at the DARK96 conference, Heidelberg (1996).

\bibitem{bionta} R.M. Bionta et al., {\sl Phys. Rev. Lett} {\bf58} 
(1987) 1494; K. Hirata et al, {\sl Phys. Rev. Lett.} {\bf 58} (1987) 
1490.

\bibitem{burrows} A. Burrows and J. Lattimer, Ap. J. {\bf 307} (1986) 178;
R. Mayle, J. Wilson and D. Schramm, Ap. J. {\bf 318} (1987) 288.

\bibitem{fr} B. Friman and O. Maxwell, Ap. J. {\bf 232} (1979) 541. For
a review, see "Stars as Laboratories for Fundamental Physics" by
G. G. Raffelt, Chicago University press (1996).

\bibitem{bhattacharya} T. Bhattacharya and P. Roy,  Phys. Rev. {\bf 
D38} (1988) 2284.

\bibitem{fukugita} M. Fukugita and N. Sakai, Phys. Lett. {\bf B114} 
(1982) 23; M. Nowakowski and S. D. Rindani, Phys. Lett. {\bf B348} 
(1995) 115.

\end{thebibliography}
\end{document}